\documentclass{appolb}
\usepackage{graphicx}
\preprint{}

\begin{document}
\bibliographystyle{num}
\title{In medium $T$ matrix with realistic nuclear 
interactions\thanks{Supported by the Polish State Committee for
Scientific Research, grant 2P03B02019.}}%
\author{Piotr Bo\.zek
\and
Piotr Czerski
\address{Institute of Nuclear Physics, 31-242 Cracow, Poland}
}
\maketitle
\begin{abstract}
We calculate the self-consistent in-medium $T$ matrix for
 symmetric nuclear matter 
using realistic interactions with many partial waves. 
 We find for the interactions used (CDBonn and Nijmegen) very
similar results for on-shell quantities. 
The effective mass and the renormalization factor
$Z_F$ at the Fermi momentum are given for a range of densities.
\end{abstract}

\thispagestyle{empty}  

The presence of a strong interaction between nucleons at short
distances
requires the use of resummation methods for many-body calculations in
nuclear matter. Besides the Brueckner-Hartree-Fock (BHF) summation of ladder
diagrams  \cite{jlm,bhf} and advanced  variational methods
\cite{vcs1,vcs2},
the self-consistent  in-medium $T$-matrix approach became feasible in
last
years \cite{di1,Bozek:1998su,Bozek:2001tz,Bozek:2002em,Dewulf:2002gi}.
The numerical method presented in \cite{Bozek:2002em} allows for
efficient calculations using off-shell nucleon propagators with
realistic interactions. Below we present results for the CDBonn
\cite{cdbonn}
and
Nijmegen  potentials \cite{nijmegen}
 incorporating all partial waves with total angular
momentum $J<9$.
The in medium $T$ matrix \cite{KadanoffBaym}
\begin{equation}
T=V+ VGGT 
\end{equation}
 is calculated with the Green's function
 $G=\left({\omega-p^2/2m -\Sigma}\right)^{-1}$ 
 self-consistently dressed
 by the self-energy in the $T$-matrix approximation
\begin{equation}
i\Sigma= Tr[T_AG]  \ .
\end{equation}
The difference with respect to the BHF approach lies in the dressing of
the
Green's functions by the
full spectral functions, which means off-shell propagation. Also,
 the double Green's function in the $T$-matrix equation
represents the propagation of two particles or two-holes unlike
in the Kernel of the Bethe-Goldstone equation in the BHF scheme where
the Pauli-blocking factor forces the propagation of two particles always.
The above equations are solved by iteration for several nuclear
densities $\rho$ in the range $0.2 \ - \ 2.4$ normal nuclear densities
$\rho_0=.16{\rm fm}^{-3}$. The details of the equations and numerics
can be found in Refs. \cite{Bozek:2001tz,Bozek:2002em}. 

We have improved the
numerical algorithm  to allow for many partial-waves in the $T$ matrix. We use
a separable parameterization of the interaction, choosing $8$ 
most important eigen-vectors of the interaction in the momentum
representation for each uncoupled partial wave and $24$ eigen-vectors for the
coupled partial waves. This parameterization is essentially equivalent to
the full parameterization in momentum.    
 In previous  $T$-matrix calculations a low-rank  separable parameterization of
the Paris potential was used. In this paper we compare the 
results for two realistic
interactions without further simplifying  approximations and taking
many partial waves. With these  calculations the binding
energy and single-particle properties can be found without
 uncertainties due to technical  simplifications.


The binding energy per particle 
in the $T$-matrix approach can be calculated from the
Koltun's sum rule
\begin{equation}
\label{bin}
E/N=\frac{1}{\rho}
\int\frac{d^3p}{(2 \pi)^3}\int_{-\infty}^{\mu}\frac{d\omega}{2\pi}
\frac{1}{2}\bigg(\frac{p^2}{2 m}+
\omega \bigg) A(p,\omega) \ 
\end{equation}
where $A$ is the nontrivial spectral function obtained for the dressed
propagators. In fact any expression for the energy should give
the same result, since the self-consistent $T$-matrix approximation is
thermodynamically consistent \cite{Baym}. The existence  a generating
function $\Phi$ for the self-energy guarantees the fulfillment of
thermodynamical relations between single-particle and global
properties of the system. The fact that the  $T$-matrix  scheme is a
consistent  (conserving) approximation has been checked explicitely in 
\cite{Bozek:2001tz} using a model interaction.

\begin{figure}
\begin{center}
\includegraphics*[width=0.75\textwidth]{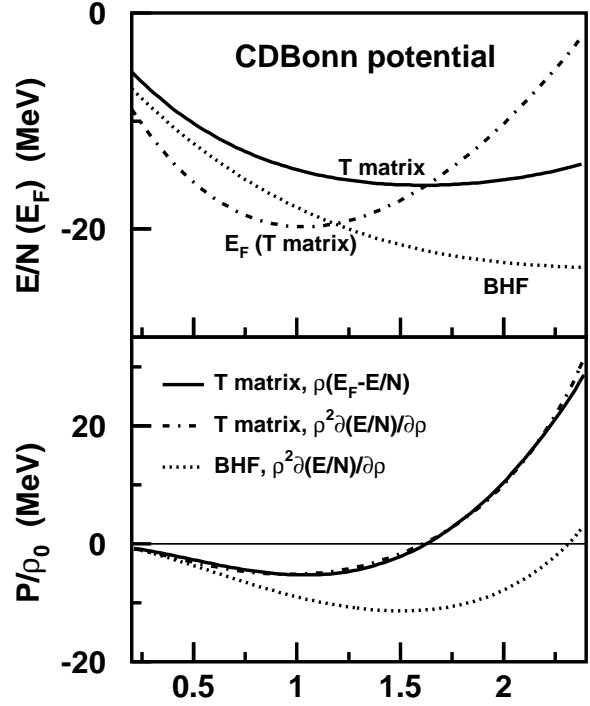}
\caption{{\bf Upper panel~:} The binding energy and the Fermi energy as a 
function of density for the $T$-matrix approach
 and the binding energy in the BHF calculation, all
for the CDBonn potential.
{\bf Lower panel~:} The pressure as a function of density
  obtained from two different expressions Eqs. (\ref{pr1})
and (\ref{pr2}). The solid and the dashed-dotted lines representing the
two results for the pressure in  the $T$-matrix calculation  lie
almost on top of each other.}
\label{becdbonn}
\end{center}
\end{figure}
\begin{figure}
\begin{center}
\includegraphics*[width=0.75\textwidth]{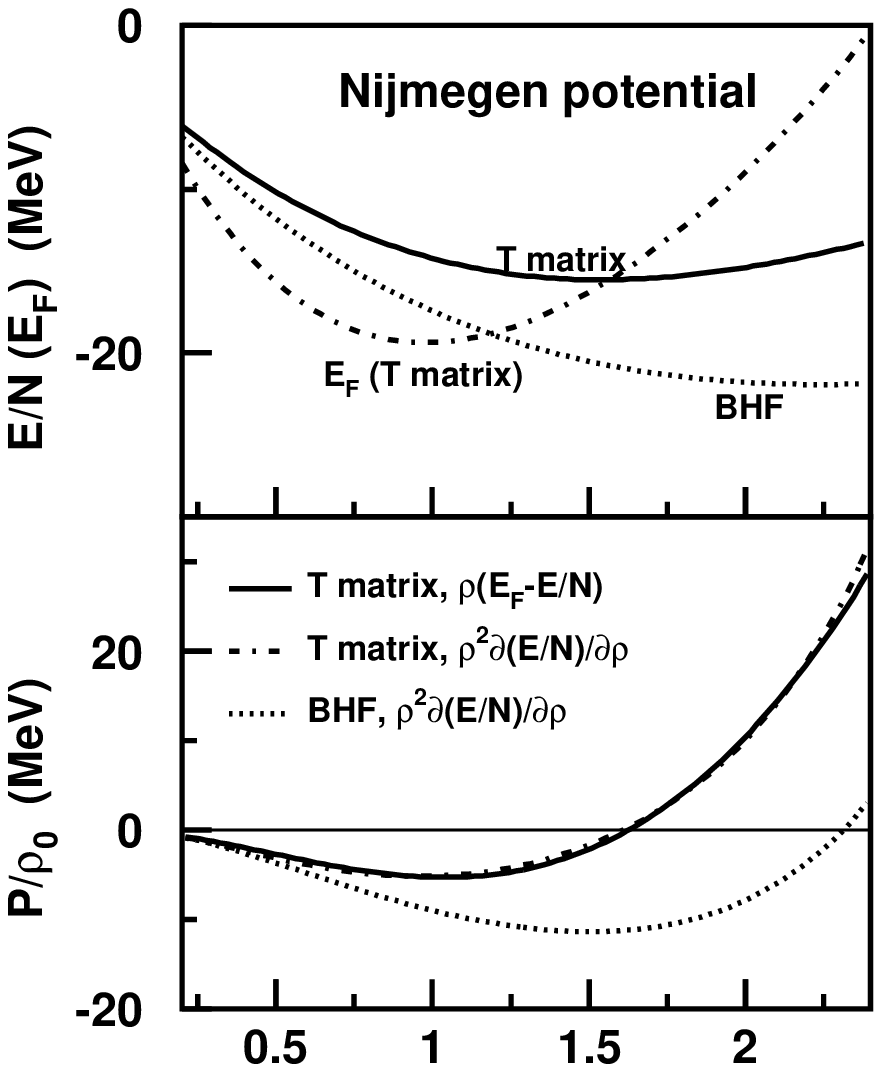}
\caption{Same as Fig. \ref{becdbonn} but for the Nijmegen potential} 
\label{benijm}
\end{center}
\end{figure}

In the upper panels of Figs. \ref{becdbonn} and \ref{benijm}
is shown the binding energy in the $T$-matrix approximation as a
function of the density in  symmetric nuclear matter compared to the
corresponding BHF results. As noted previously the $T$ matrix gives
smaller binding energies and smaller saturation densities than the BHF
calculation. At low densities the BHF and the $T$-matrix results
converge as expected. The  interactions studied in this work give
similar results, e.g at normal nuclear  density
 $E/N=-14.3$ and $-14.1$MeV for the CDBonn and
Nijmegen interaction respectively. The corresponding BHF binding
energies at that density are $3.3-3.5$MeV lower. In the same figures the
Fermi energy $E_F$ is shown. The Fermi
energy in the $T$-matrix calculation is consistent with the binding
energy; the Hugenholz-Van Hove relation \cite{hvh}
\begin{equation}
E_F=E/N \ \ {\rm (~at~saturation~density~)}
\end{equation}
is automatically fulfilled.

The pressure in the system can be obtained from several equivalent
expressions.
We consider 
\begin{equation}
\label{pr1}
P=\rho^2\frac{\partial(E/N)}{\partial \rho}
\end{equation}
and
\begin{equation}
\label{pr2}
P=\rho\left(E_F- E/N\right) \ .
\end{equation}
From the second form follows the Hugenholz-Van Hove relation at the
saturation point ($P=0$).
The  expressions (\ref{pr1}) and (\ref{pr2}) are equivalent in the
self-consistent $T$-matrix approximation.
In the BHF approach
 the pressure can be calculated as the derivative of the binding energy 
(\ref{pr1}).

\begin{figure}
\begin{center}
\includegraphics[width=0.75\textwidth]{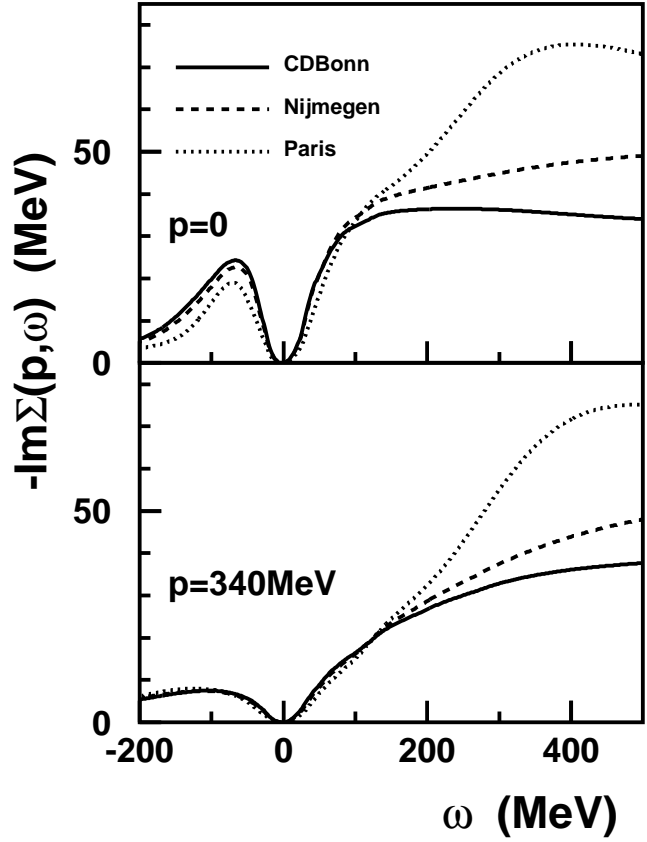}
\caption{The imaginary part of the retarded self-energy 
$-{\rm Im}\Sigma(p,\omega)$ as a function of the energy $\omega$ for
  $p=0$ (upper panel) and $p=340$MeV (lower panel) for different 
nucleon-nucleon potentials.}
\label{gamo}
\end{center}
\end{figure}
One-body properties are determined by the self-energy.
In Fig. \ref{gamo}
is shown the imaginary part  of the self-energy off-shell 
$-{\rm Im}\Sigma(p,\omega)$. The self-energy is very similar for the CDBonn
and Nijmegen potentials and close to the result for the Paris potential
\cite{Bozek:2002em} for energies close to the Fermi energy
($|\omega|<200$MeV)
and for momenta up to $500$MeV. 
At higher energies differences start to
appear, because of different off-shell behavior of the $T$-matrix for
different interactions.
The Paris interaction we use is a separable parameterization
with a small number of partial waves, this  can lead to additional
differences as compared to  results from the other two potentials.

\begin{figure}
\begin{center}
\includegraphics[width=0.75\textwidth]{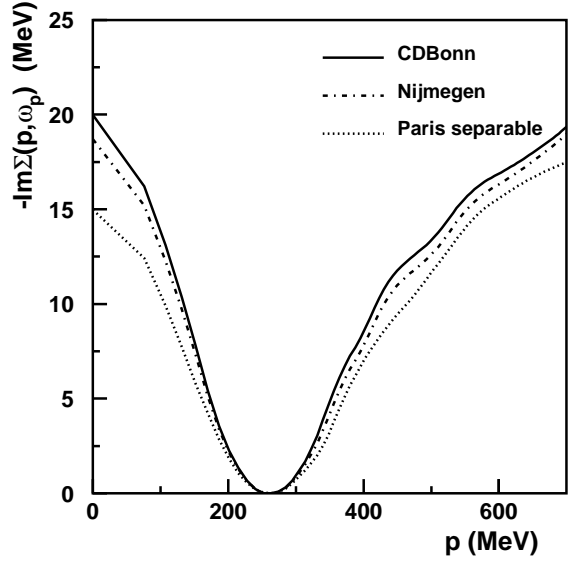}
\caption{The imaginary part of the retarded self-energy 
$-{\rm Im}\Sigma(p,\omega_p)$ at the quasiparticle pole 
as a function of the momentum.}
\label{selfi}
\end{center}
\end{figure}
As can be seen from Fig. \ref{selfi} the self-energy on-shell, \ie for
$\omega=\omega_p=p^2/2m+{\rm Re}\Sigma(p,\omega_p)$,
 is similar for different interactions.
For momenta $p<700$MeV the width of the quasiparticle excitation is
similar. It is always zero at the Fermi momentum and increases
quadratically when going away from it. The quadratic increase of
the single-particle width is given by the cross section and the density of
states at the Fermi surface. The cross sections are similar for
different parameterizations of the nuclear potentials and the density
of states at the Fermi surface is determined by the effective mass and
the renormalization factor 
\begin{equation}
Z_p=\left(1-\partial {\rm
  Re}\Sigma(p,\omega)/\partial \omega\right)|_{\omega=\omega_p} \ .
\label{zfeq}
\end{equation}
 As shown below, those
quantities obtained  for different nucleon-nucleon
interactions are similar.

\begin{figure}
\begin{center}
\includegraphics[width=0.75\textwidth]{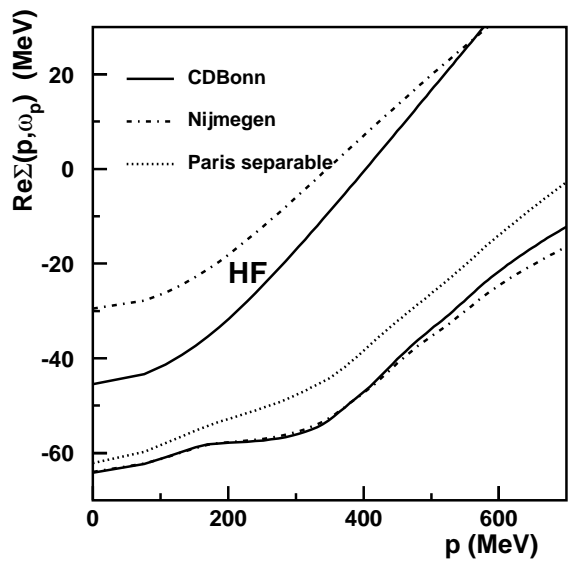}
\caption{The single-particle potential ${\rm Re}\Sigma(p,\omega_p)$ 
at the quasiparticle pole 
as a function of the momentum. The upper curves represent the
Hartree-Fock contribution for the Nijmegen and CDBonn interactions.}
\label{selfr}
\end{center}
\end{figure}
The real part of the self-energy determines the single-particle
potential, \ie the real part of the optical potential.
The fulfillment of the Hugenholz-Van Hove relation
guarantees the correct normalization of the single-particle potential.
The real part of the self-energy is given as the sum of the
Hartree-Fock and   dispersive contributions
\begin{equation}
{\rm Re}\Sigma(p,\omega)=\Sigma_{HF}(p)+\int\frac{d
  \omega^{'}}{\pi}\frac{{\rm
    Im}\Sigma(p,\omega^{'})}{\omega-\omega^{'}} \ .
\label{opteq}
\end{equation}
 The Hartree-Fock part is different for the Nijmegen and CDBonn
potentials, but
the total single-particle energy on-shell is very similar
(Fig. \ref{selfr}). The difference in the Hartree-Fock part is
compensated by the dispersive part, which  includes integration of
the imaginary part of the self-energy far off-shell. The differences in
${\rm Im}\Sigma(p,\omega)$ for $\omega>200$MeV (Fig. \ref{gamo})
give the necessary
shifts leading to similar total optical potentials (\ref{opteq}).
Although the Nijmegen and CDBonn interactions show some differences in
the off-shell behavior of the self-energies, the resulting
properties of the quasiparticle pole are very similar in the range of
densities $\rho<2\rho_0$.

\begin{figure}
\begin{center}
\includegraphics[width=0.75\textwidth]{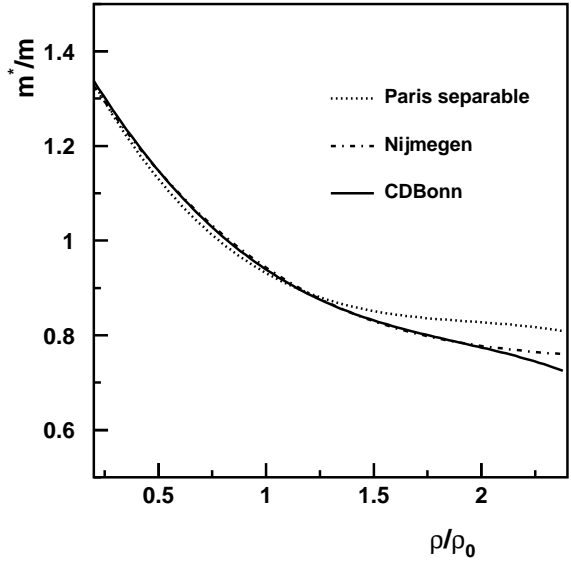}
\caption{The effective mass $m^\star/m$ at the Fermi surface for
  different nuclear interactions.}\label{meff}
\end{center}
\end{figure}
The properties of the quasiparticles at the Fermi surface are defined
by the renormalization factor (\ref{zfeq}) and the effective mass
\begin{equation}
m^\star=\frac{pd p}{d \omega_p} \ .
\end{equation}
The effective mass and the renormalization factor are important
to define the effective interaction between quasiparticles 
\cite{Dickhoff:1999yi,Schwenk:2002fq} and  the superfluid gap
\cite{Bozek:2000fn,Bozek:2002jw}.
We present the calculation of these quantities for different
interactions in a range of densities.
\begin{figure}
\begin{center}
\includegraphics[width=0.75\textwidth]{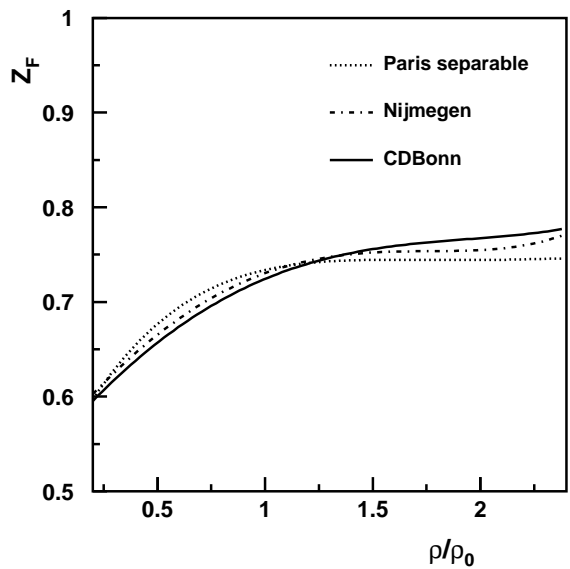}
\caption{The renormalization factor of the quasiparticle pole  $Z_p$
at the Fermi surface for
  different nuclear interactions.}\label{zeff}
\end{center}
\end{figure}
We find that for $\rho<2 \rho_0$ the CDBonn  and the 
Nijmegen interactions give very similar results
 (Figs. \ref{meff}, \ref{zeff}).
 The difference of the
calculation using the separable Paris potential can be attributed to the
simplicity of that parameterization.
The effective mass in our calculation comes out very close to the free
nucleon mass. It is useful for further applications to parameterize its
density dependence 
\begin{equation}
\frac{m^\star}{m}=1.43 -.623\frac{\rho}{\rho_0}+.146
\left(\frac{\rho}{\rho_0}\right)^2 
\end{equation}
and similarly the for renormalization $Z_p$ at the Fermi  momentum
(Fig. \ref{zeff})
\begin{equation}
Z_{F}=.57+.2\frac{\rho}{\rho_0}-.05
\left(\frac{\rho}{\rho_0}\right)^2 \ .
\end{equation}
At normal nuclear density $Z_F\simeq0.72$ which means a reduction of
the effective interaction between quasiparticles by a factor
$Z_F^2\simeq0.5$.

We  extend previous calculations of the self-consistent
$T$ matrix to soft core CDBonn and Nijmegen nucleon-nucleon
potentials.
We include in the calculation partial waves up to the total angular
momentum $J=8$. We present for the first time results on the binding energy
for such realistic and detailed parameterizations of the two-body
interactions. The binding energy leads to a harder equation of state,
a smaller binding energy and a smaller saturation density than the BHF 
approximation. The pressure we find is thermodynamically consistent as
expected for a conserving approximation.
The single-particle energy and the width at the quasiparticle pole
come out similarly for different interactions used.
It is remarkable that differences in the Hartree-Fock energies and 
differences in the 
off-shell behavior of the imaginary part of the self-energy cancel
out in the single-particle potential.
We present results on the properties of the quasiparticle pole at
the Fermi surface.
The effective mass is close to the free one and the renormalization
factor of the quasiparticle pole is $Z_F\simeq0.7$
around the normal nuclear density.

\bibliography{../mojbib}

\begin{thebibliography}{10}
\expandafter\ifx\csname url\endcsname\relax
  \def\url#1{\texttt{#1}}\fi
\expandafter\ifx\csname urlprefix\endcsname\relax\def\urlprefix{URL }\fi

\bibitem{jlm}
J.~P. Jeukenne, A.~Leugeunne, C.~Mahaux, Phys. Rep., {\bf 25} (1976) 83.

\bibitem{bhf}
R.~Brockmann, R.~Machleit, Phys. Rev., {\bf C42} (1990) 1965.

\bibitem{vcs1}
R.~B. Wiringa, V.~Fiks, A.~Fabrocini, Phys. Rev. C, {\bf 38} (1988) 1010.

\bibitem{vcs2}
A.~Akmal, V.~R. Pandharipande, D.~G. Ravenhall, Phys. Rev. C, {\bf 58} (1998)
  1804.

\bibitem{di1}
W.~H. Dickhoff, Phys. Rev., {\bf C58} (1998) 2807.

\bibitem{Bozek:1998su}
P.~Bo\.zek, Phys. Rev., {\bf C59} (1999) 2619.

\bibitem{Bozek:2001tz}
P.~Bo\.zek, P.~Czerski, Eur. Phys. J., {\bf A11} (2001) 271.

\bibitem{Bozek:2002em}
P.~Bo\.zek, Phys. Rev., {\bf C65} (2002) 054306.

\bibitem{Dewulf:2002gi}
Y.~Dewulf, D.~Van~Neck, M.~Waroquier, Phys. Rev., {\bf C65} (2002) 054316.

\bibitem{cdbonn}
R.~Machleidt, F.~Sammatrruca, Y.~Song, Phys. Rev., {\bf C53} (1996) R1483.

\bibitem{nijmegen}
V.~G. Stoks, R.~A.~M. Klomp, M.~C.~M. Rentmeester, J.~J. de~Swart, Phys. Rev.,
  {\bf C48} (1993) 792.

\bibitem{KadanoffBaym}
L.~Kadanoff, G.~Baym, {\em Quantum Statistical Mechanics}, Bejamin, New York,
  1962.

\bibitem{Baym}
G.~Baym, Phys. Rev., {\bf 127} (1962) 1392.

\bibitem{hvh}
N.~Hugenholz, L.~V. Hove, Physica, {\bf 24} (1958) 363.

\bibitem{Dickhoff:1999yi}
W.~H. Dickhoff, C.~C. Gearhart, E.~P. Roth, A.~Polls, A.~Ramos, Phys. Rev.,
  {\bf C60} (1999) 064319.

\bibitem{Schwenk:2002fq}
A.~Schwenk, B.~Friman, G.~E. Brown, Nucl. Phys., {\bf A713} (2003) 191--216.

\bibitem{Bozek:2000fn}
P.~Bo\.zek, Phys. Rev., {\bf C62} (2000) 054316.

\bibitem{Bozek:2002jw}
P.~Bo\.zek, Phys. Lett., {\bf B551} (2002) 93.

\end{thebibliography}

\end{document}